\documentclass[traditabstract]{aa} 

\usepackage{graphicx}
\usepackage{natbib}
\usepackage{txfonts}

\newcommand{\lya}{Ly$\alpha$ }

\begin{document}

   \title{Beamed Ly$\alpha$ Emission through Outflow-Driven Cavities}

   \author{C. Behrens
          \inst{1}
          \and
         M. Dijkstra\inst{1,2,3}
         \and 
         J.C. Niemeyer\inst{1}
         \fnmsep
          }
   \institute{Institut f\"ur Astrophysik, Georg-August Universit\"at G\"ottingen,
              Friedrich-Hund-Platz 1, D-37077 G\"ottingen\\
              \email{cbehren@astro.physik.uni-goettingen.de/niemeyer@astro.physik.uni-goettingen.de}
	\and
	    Max Planck Institute for Astrophysics, Karl-Schwarzschild-Str. 1, 85741, Garching, Germany\\
	\and
	    Institute of Theoretical Astrophysics, University of Oslo, Postboks 1029, 0858Oslo, Norway\\
	     \email{mark.dijkstra@astro.uio.no}}
   \date{Draft \today}

 
  \abstract{
We investigate the radiative transfer of Ly$\alpha$ photons through simplified anisotropic gas distributions, which represent physically motivated extensions of the popular 'shell-models'. Our study is motivated by the notion that ({\it i}) shell models do not always reproduce observed Ly$\alpha$ spectral line profiles, ({\it ii}) (typical) shell models do not allow for the escape of ionizing photons, and ({\it iii}) the observation \& expectation that winds are more complex, anisotropic phenomena. We examine the influence of inclination on the Ly$\alpha$ spectra, relative fluxes and escape fractions.  We find the flux to be enhanced/suppressed by factors up to a few depending on the parameter range of the models, corresponding to a boost in equivalent width of the same amplitude if we neglect dust. In general, lower mean optical depths tend to reduce the impact of anisotropies as is expected. We find a correlation between an observed peak in the -- occasionally triple-peaked -- spectrum at the systemic 
velocity and the existence of a low optical depth cavity along the line of sight. This can be of importance in the search for ionizing photons leaking from high-
redshift galaxies since these 
photons will also be able to escape through the cavity.}

   \maketitle
%

\section{Introduction}

There is ample observational evidence that Ly$\alpha$ photons scatter through outflowing HI gas as they emerge from the interstellar medium of star forming galaxies. In local galaxies, gas kinematics on interstellar scales is one of the key factors that determines whether Ly$\alpha$ photons escape \citep[][]{Kunth98,Atek08}. Moreover, observed Ly$\alpha$ spectral lines in higher redshift ($z\sim 3$) galaxies have P-Cygni type profiles which consist of a redshifted emission line and a blueshifted absorption trough, which are clear characteristics of the presence of an outflow \citep[e.g.][]{Shapley03}. The ubiquitous presence of outflows is also revealed by the blueshifted low-ionization metal absorption lines in a large fraction of Lyman Break Galaxies (LBGs).  These same LBGs typically exhibit a Ly$\alpha$ emission line that is redshifted by $\sim 2-3\hspace{1mm} \times$ the outflow velocity inferred from the absorption line data, and strongly suggest that Ly$\alpha$ photons have been scattered by \ion{H}{I}
 within the outflow \citep[e.g.][]{Steidel10}. 

Modeling cold, outflowing interstellar gas is an extremely difficult task, and practically requires simulations to fully resolve the interstellar medium \citep[see][and references therein]{DK12}. Models of Ly$\alpha$ emitting galaxies have thus far employed a `subgrid' model of scattering through the outflow, in which the outflow is represented by a spherical shell of HI gas \citep[see][for a description of the shell model]{Ahn03,Verhamme06}. This spherical shell is geometrically thin (thickness is 10\% of the radius of the shell), and is parameterized (primarily) by an HI column density, an outflow velocity, and a dust optical depth. It has been demonstrated that these models can provide good fits to observed Ly$\alpha$ line shapes \citep{Verhamme08,Schaerer08,Vanzella10}. 

Recently however, NIR spectrographs have measured rest-frame optical nebular lines from galaxies at $z\sim 2$ (where it is possible to detect H$\alpha$) and $z\sim 3$ (where it is possible to detect [OIII]). The detection of these nebular lines constrains the systemic velocity of galaxies, and therefore the spectral shift of the Ly$\alpha$ line relative to their systemic velocity \citep{McLinden11,Kulas12,Hashimoto13}. Importantly, it has been demonstrated that having both the shift {\it and} the spectral line shape poses a challenge to the shell models for a fraction of Ly$\alpha$ emitting sources, i.e. since the Doppler parameter $b$ derived from this data seems not to fit the observed low-ionization absorption \citep{Kulas12}, although the exact fraction of emitters problematic for the shell model is still debatable. There is also evidence for the existence of holes in the HI distribution of some LAEs that are only poorly fitted by the shell model \citep{Chonis13}. Hence, there is increasing observational demand for models that go beyond the shell model. 
Moreover, from a physical point of view the shell-model is not satisfactory
because its simple geometry does not capture the complex structure of the gas in outflows.

Going beyond the shell model is clearly complicated: breaking spherical symmetry causes the predicted spectral line shapes to depend on viewing angle. Moreover, departures from spherical symmetry introduce new parameters that significantly increase the volume of parameter space that can be explored. In this paper, we investigate Ly$\alpha$ transfer through non-spherical models for winds that are still relatively simple \& physically motivated. Examples of models we study include: ({\it i}) shells with holes, ({\it ii}) bipolar winds, and ({\it iii}) so called `cavity' models (see \S~\ref{sec:bi}-\S~\ref{sec:cavity} for details). The first two classes of models approach a spherical shell model when either the hole becomes arbitrarily small, or when the anisotropic velocity component of the bipolar model vanishes. 
\citet{Duval2013} have investigated different modifications of the shell model, namely statistically isotropic, dusty and clumpy shells. 
These models are in statistical sense still isotropic and have covering factors near 1, yielding only few lines of sight that are optically thin, while the models discussed here have strong anisotropies with many optically thin lines of sight. Importantly, \citet{Duval2013} (and \citet{Laursen13}) have focussed on studying boosts in EW, averaged over all sightlines. \citet{Laursen13} points out that there can be departures in this boost from the average along individual sightlines, which is the main focus of our work.

A common property of all our models is that there exist sightlines through the wind which contain a very low (or zero) HI-column density. Our wind models thus include possible escape routes for ionizing photons (which clearly do not exist in the shell models). This property of our models is important, and motivated by a scenario in which ionizing photons escape efficiently only along a fraction of sightlines. This picture is motivated by observations of LBGs for which outflowing low-ionization absorption line systems, which trace `cold' ($T\sim 10^4$ K) gas, are not always fully covering the UV-emitting regions\footnote{A caveat is that the covering factor is less than unity over a range of velocities. Cold gas moving at different velocities can cover different parts of unresolved UV emitting sources, and it remains possible that the covering factor integrated over velocity is 1.}; For example, Jones et al. (2013) find that the maximum low-ionization covering fraction for their $z>2$ is $100\%$ in only 
2 out of 8 galaxies (also see Heckman et al. 2011, who find evidence for a low covering factor of optically thick, neutral gas in a small fraction of lower redshift Lyman Break Analogues). This covering factor of cold outflowing gas decreases with redshift \citep{Jones12,Jones13}, which may explain that the escape fraction of ionizing photons increases with redshift as has been inferred\footnote{This redshift evolution appears to be required in order to be able to reionize the Universe \citep[][]{Kuhlen12}.} independently by observations of the Ly$\alpha$ forest and the UV-LF of drop-out galaxies  \citep[e.g.][]{Kuhlen12,Robertson13}. Finally, the possibility that ionizing photons escape highly anisotropically naturally explains the apparently bimodal distribution of the inferred escape fraction in star forming galaxies, in which a small fraction of galaxies have a large $f_{\rm esc}$ and a large fraction practically are consistent with having $f_{\rm esc}=0$ \citep[][]{Shapley06,Nestor11,Vanzella12}.

From a Ly$\alpha$ transfer point-of-view, the existence of low HI-column density channels can have major observational consequences. For non-spherical gas distribution, Ly$\alpha$ generally escapes anisotropically \citep[see][and references therein]{Laursen07,Verhamme12,Zheng13}, where the flux can vary by as much as a factor of $>10$ depending on the viewing angle \citep{Verhamme12}. We expect this anisotropy to be potentially more extreme in models that contain low-N$_{\rm HI}$ sightlines, which we refer to as `holes' and/or `cavities'. These cavities may correspond to regions that have been cleared of gas and dust by feedback processes \citep[see][who describe a simple `blow-out' model]{Nestor11,Nestor13}.

In this paper we study Ly$\alpha$ transfer through simplified models of the ISM which include outflow-driven cavities. Our models consist of several `families'. The `shells-with-holes' and `bipolar-wind' models can easily be connected to the traditional shell models, which makes it easier to interpret our results. The `cavity' models represent a scenario in which gas is outflowing perpendicular to a galactic disk. This paper has two main goals: ({\it i}) to investigate the variation in flux for different viewing angles which will give us an estimate for the expected boost in equivalent width (EW) with respect to the isotropic case, and ({\it ii}) to investigate whether there are robust observational spectral signatures of the Ly$\alpha$ line for sightlines that go through a cavity.

The outline of this paper is as follows: in \S~\ref{sec:models} we introduce our models. In \S~\ref{sec:individual}, we discuss and interpret our results with a focus on individual dust-free realizations of our model families. In  \ref{sec:dust} and \S~\ref{sec:pstudy}, we consider the influence of dust and generalize our results with a parameter study.

\section{Our Models and Parameters}
\label{sec:models}

In order to investigate the variation of fluxes and spectra with viewing angle, we have to break the symmetry of the models by introducing anisotropic velocity and/or density fields. Like the shell model, our models are primarily motivated by situations where galactic winds and other feedback mechanisms drive outflows, blowing out or ionizing neutral gas. We take into the account the asymmetry of such winds that are expected to occur particularly in the transverse direction of the plane. This introduces an anisotropy in this direction. We define the `inclination' as the angle between the z-axis and a unit vector towards the observer. To quantify the anisotropy of fluxes and spectra, we define a set of observers, each observing photons within a certain angular bin. Bins are equally spaced in $\cos(\theta)$ so that each observer sees the same area of the unit sphere, meaning that in isotropic cases, all observers should detect an equal number of photons.

\subsection{The 'Bipolar' Model}
\label{sec:bi}
   \begin{figure*}
   \centering
   \includegraphics[trim=20cm 15cm 0cm 10cm,clip=true,width=6cm]{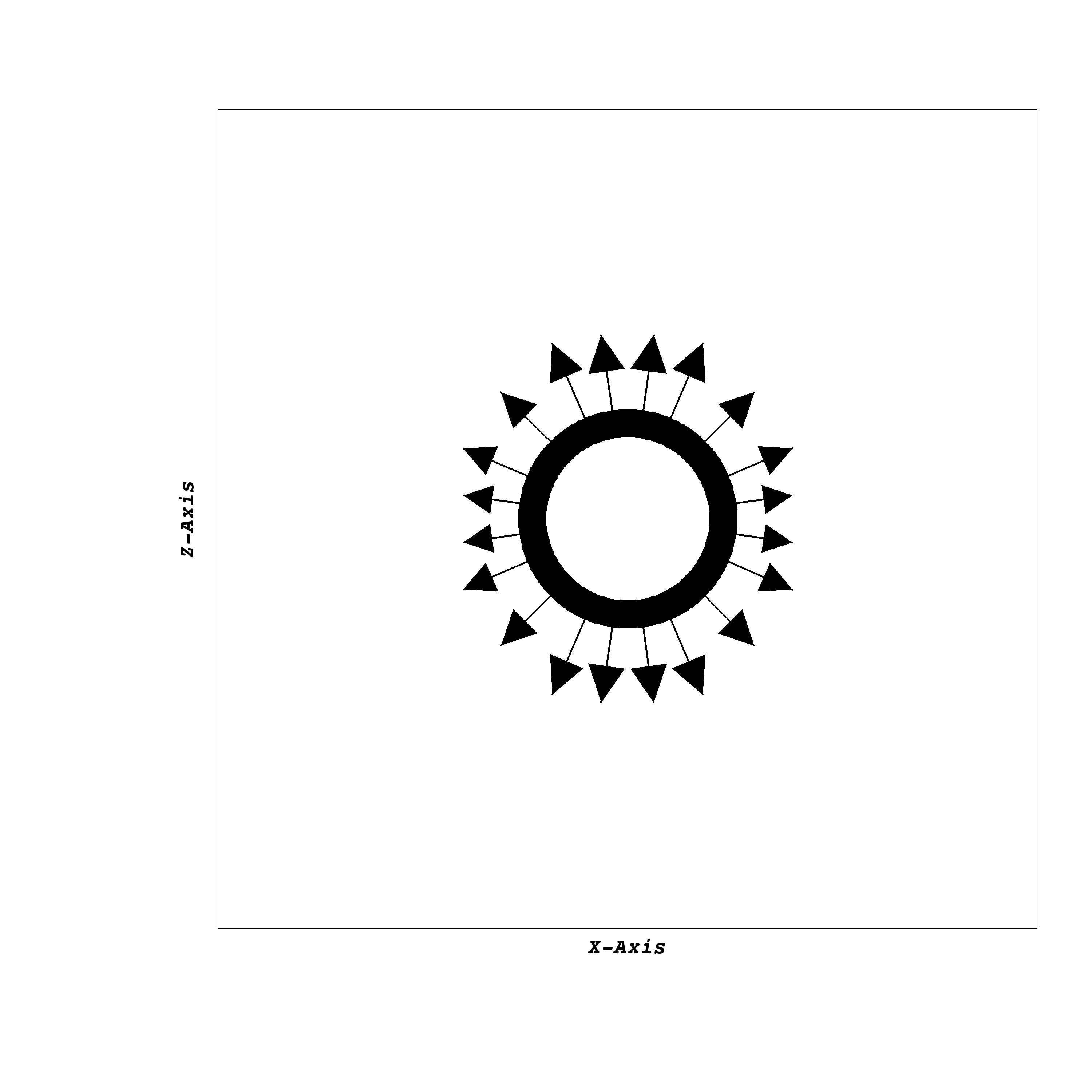}
   \includegraphics[trim=20cm 15cm 0cm 10cm,clip=true,width=6cm]{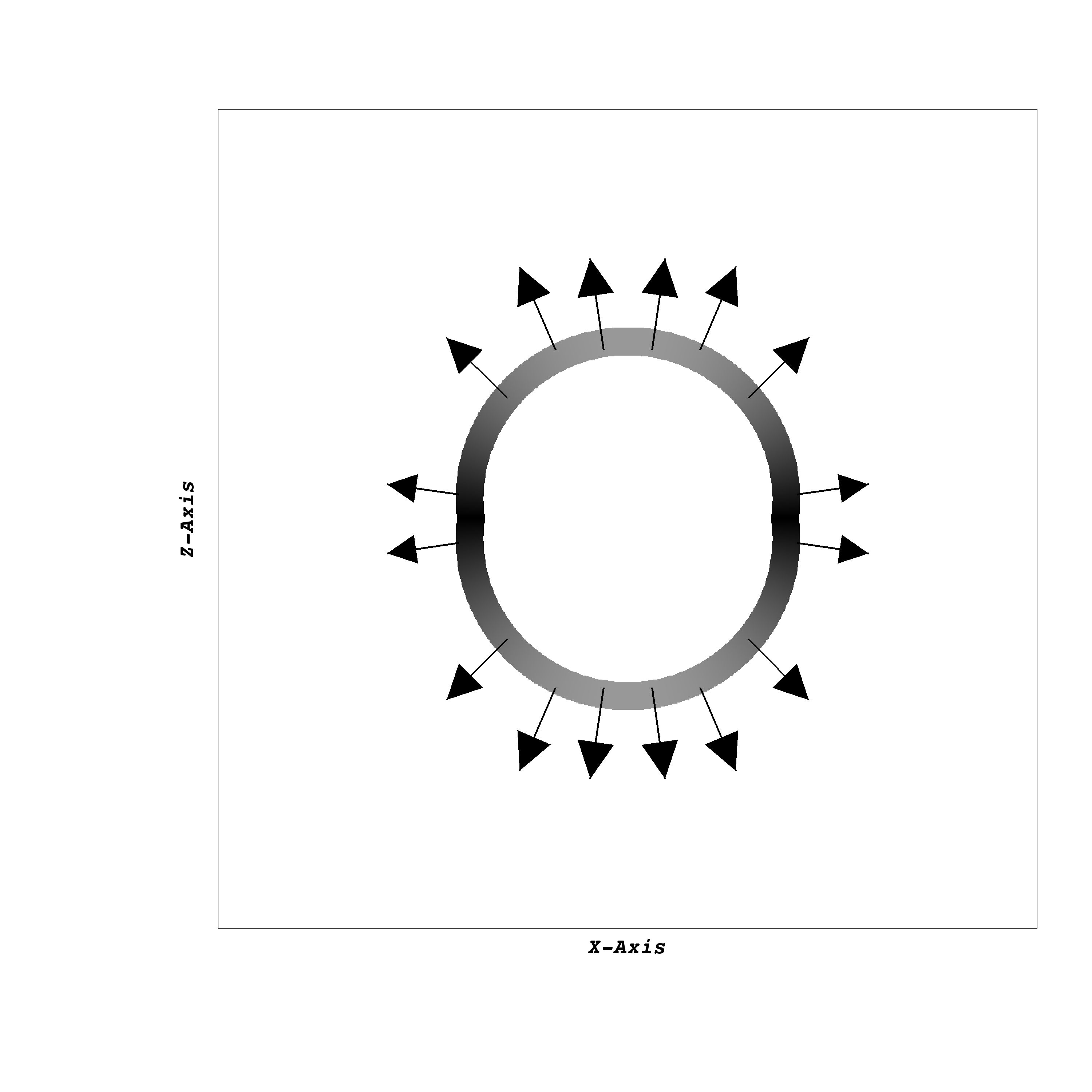}
   \includegraphics[trim=20cm 15cm 0cm 10cm,clip=true,width=6cm]{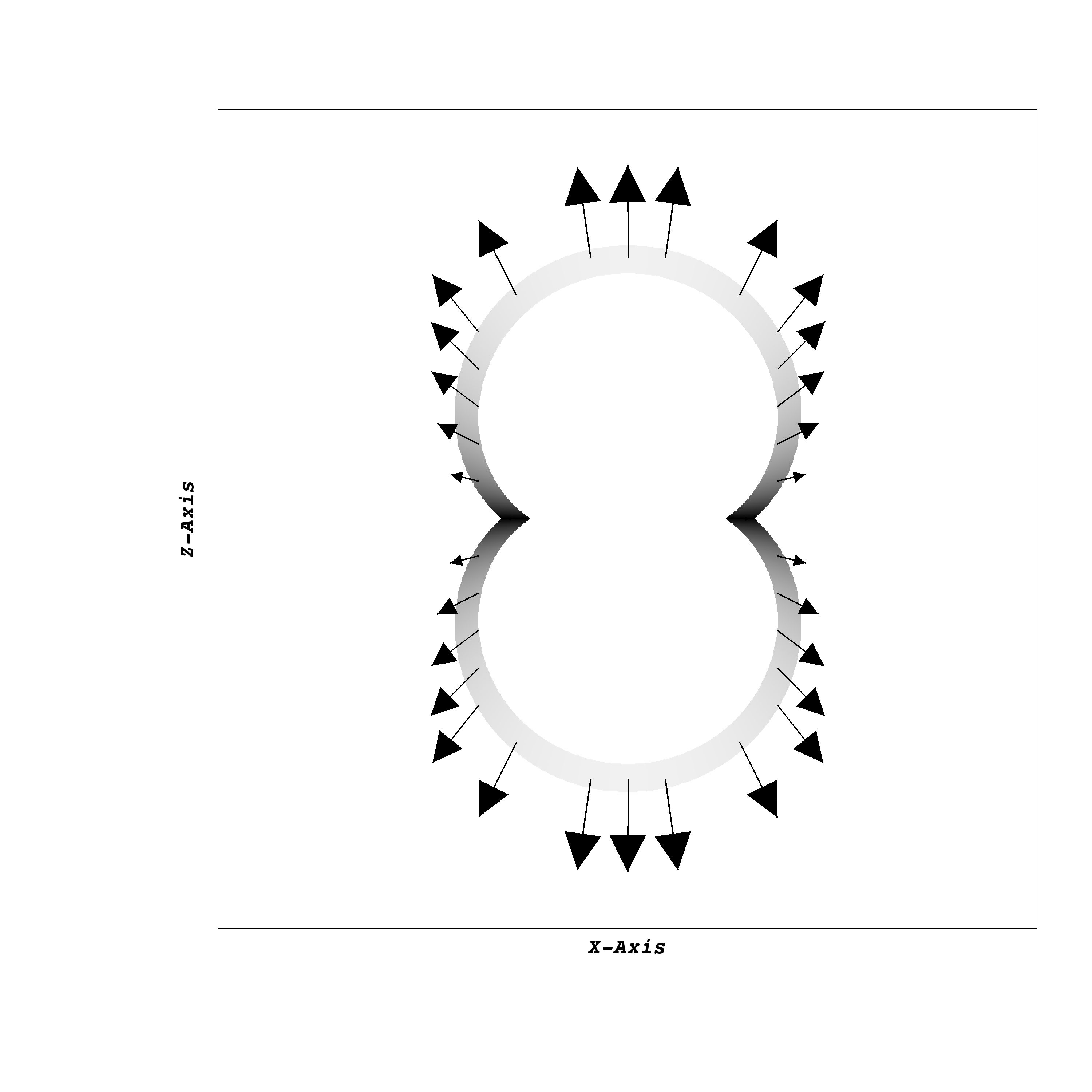}
      \caption{Illustration of the 'Bipolar' model for three different time parameters. Arrows indicate velocity vectors, contours the density distribution. As the time parameter evolves from left to right, the anisotropy in the bulk velocity becomes visible in the density distribution.}
         \label{fig:setup_bipolar}
   \end{figure*}

Here, we assume the wind to drive the expanding shell asymmetrically. This deforms the shell over time and displaces it anisotropically. Likewise, the density evolves with inclination, because mass conservation ensures that the mass within a solid angle remains constant. We define the velocity at an angle $\theta$ to be
\begin{equation}
 \vec v(\vec r) = \vec e_r (v_b \cos(\theta)+ v_c)\,\,,
\end{equation}
where $\vec e_r$ is the unit vector from the center to $\vec r$, $v_b$ is the magnitude of the anisotropic velocity component and $v_c$ is the isotropic component. At a given inclination $\cos \theta$, the shell will be displaced by some distance $s$ from the original position,
\begin{equation}
 s = (v_b \cos(\theta)+ v_c)t\,.
\end{equation}
Here, $t$ is some time parameter that controls the evolution of the shell's shape. We caution the reader that for our radiative transfer, we assume a steady state, e.g. $t$ does not evolve during the calculation. $t$ is given in units of the simulation box crossing time, e.g. $s$ becomes the box length for $t=1$.
The density within the shell at a given inclination reads
\begin{equation}
 \rho = \rho_0 \frac{r_i^2 + r_i r_o + r_o^2}{3 s^2 + r_i^2 + r_i r_o + r_o^2 + 3 s (r_i + r_o)}\,\,,
\end{equation}
where $\rho_0$ is some hydrogen density defined such that at $t=0$
\begin{equation}
 N^{'}_H = \rho_0 (r_o - r_i)\,\,.
\end{equation}
We denote this column density with an additional prime because it is not the effective column density if $t\neq0$.
In Fig. \ref{fig:setup_bipolar}, this setup and its dependence on the time parameter is illustrated. In practice, we parametrize $v_c$ and $v_b$ by 
the sum of both $v_{tot}$ and the fraction $f_b$ that goes in to 
parameterizing the anisotropic velocity field
\begin{equation}
 v_b = f_b v_{tot}\,\,.
\end{equation}
\subsection{The 'Shell with Holes' Model}
\label{sec:hole}

   \begin{figure}[!h]
   \centering
   \includegraphics[trim=0cm 0cm 0cm 0cm,clip=true,width=0.7\linewidth]{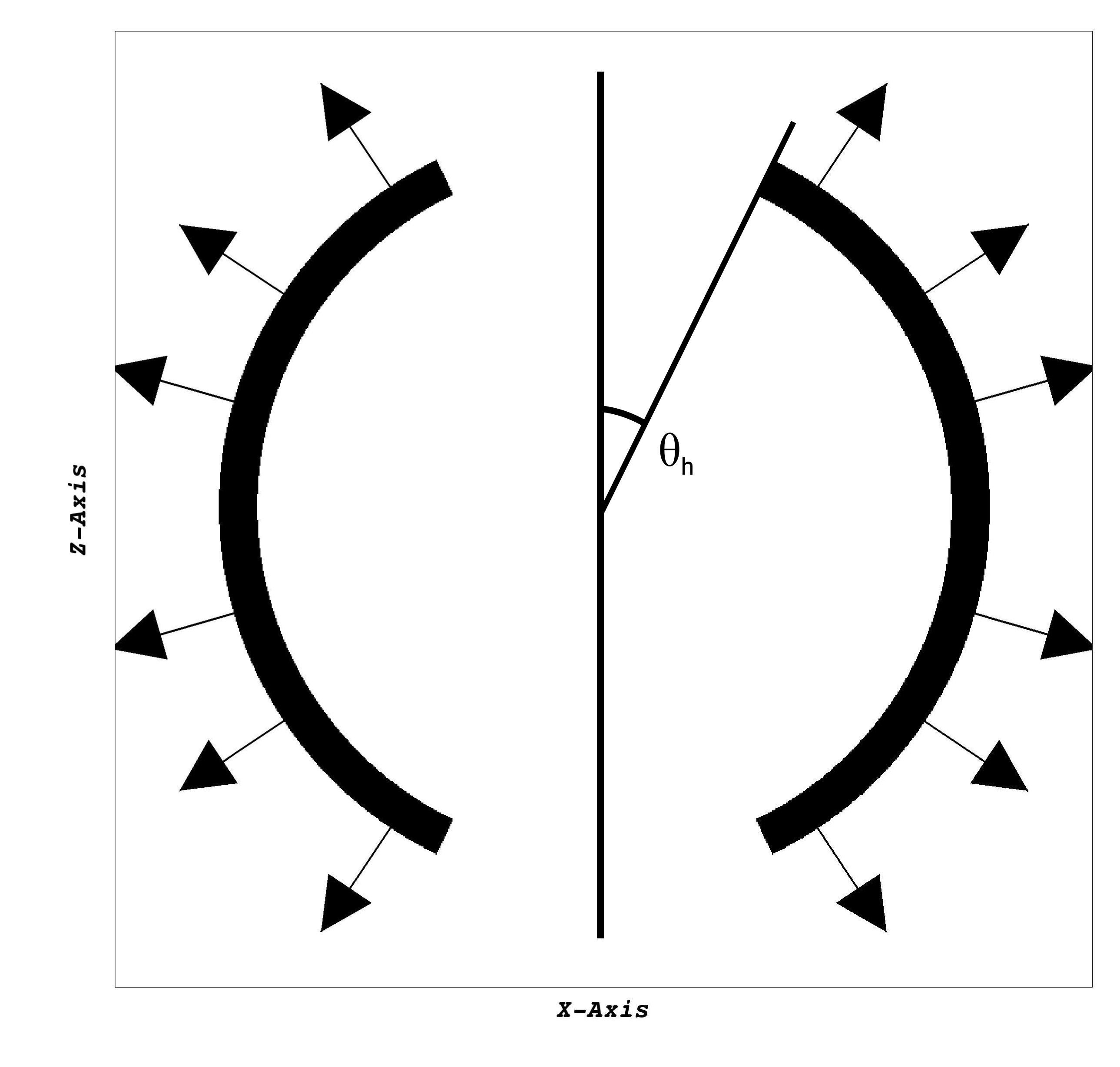}
      \caption{Illustration of the 'Shell with Holes' model. Arrows indicate velocity vectors, contours the density distribution.}
         \label{fig:setup_holes}
   \end{figure}

Assuming a spherical expanding hydrogen shell around the emitter, in one limiting case the wind emerging from the ISM might just blow off the cap of the spherical shell in the $\pm z$-direction, carving a conical region out of the shell. A cut through the density field of such a setup can be seen in Fig. \ref{fig:setup_holes}. The solid angle subtended by the conical holes in the shell, $\Omega$, is the only new parameter in this setup and can be also parametrized by the angle $\theta_H$ as indicated in Fig. \ref{fig:setup_holes}. As usual, we measure the frequency of escaping photons in the dimensionless quantity $x=\Delta \nu/\nu_D$, where $\Delta \nu$ is the physical frequency centered on the \lya line center and $\nu_D$ is the Doppler frequency.
\subsection{The 'Cavity' Model}
\label{sec:cavity}
   \begin{figure*}
   \centering
   \includegraphics[trim=20cm 15cm 0cm 10cm,clip=true,width=6cm]{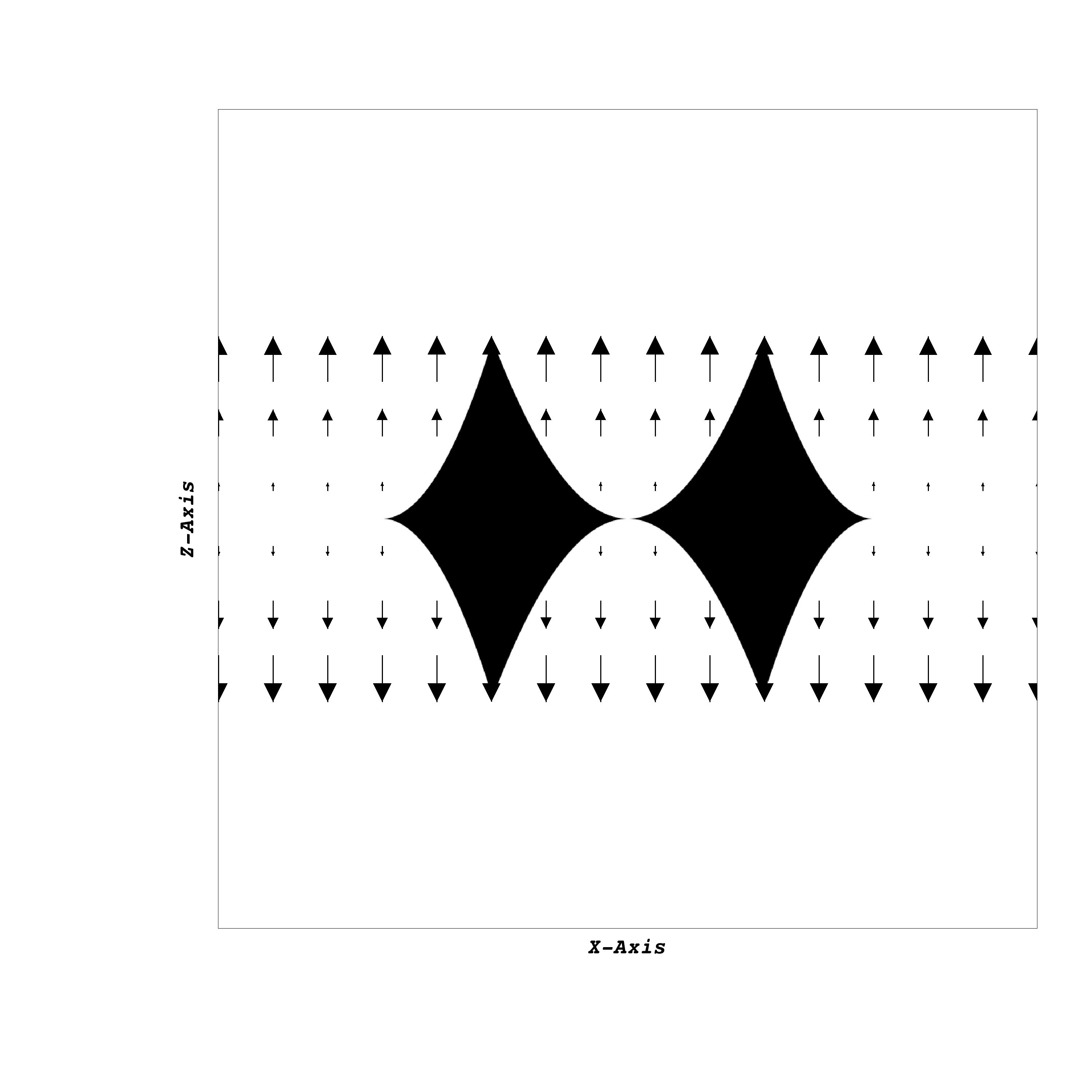}
   \includegraphics[trim=20cm 15cm 0cm 10cm,clip=true,width=6cm]{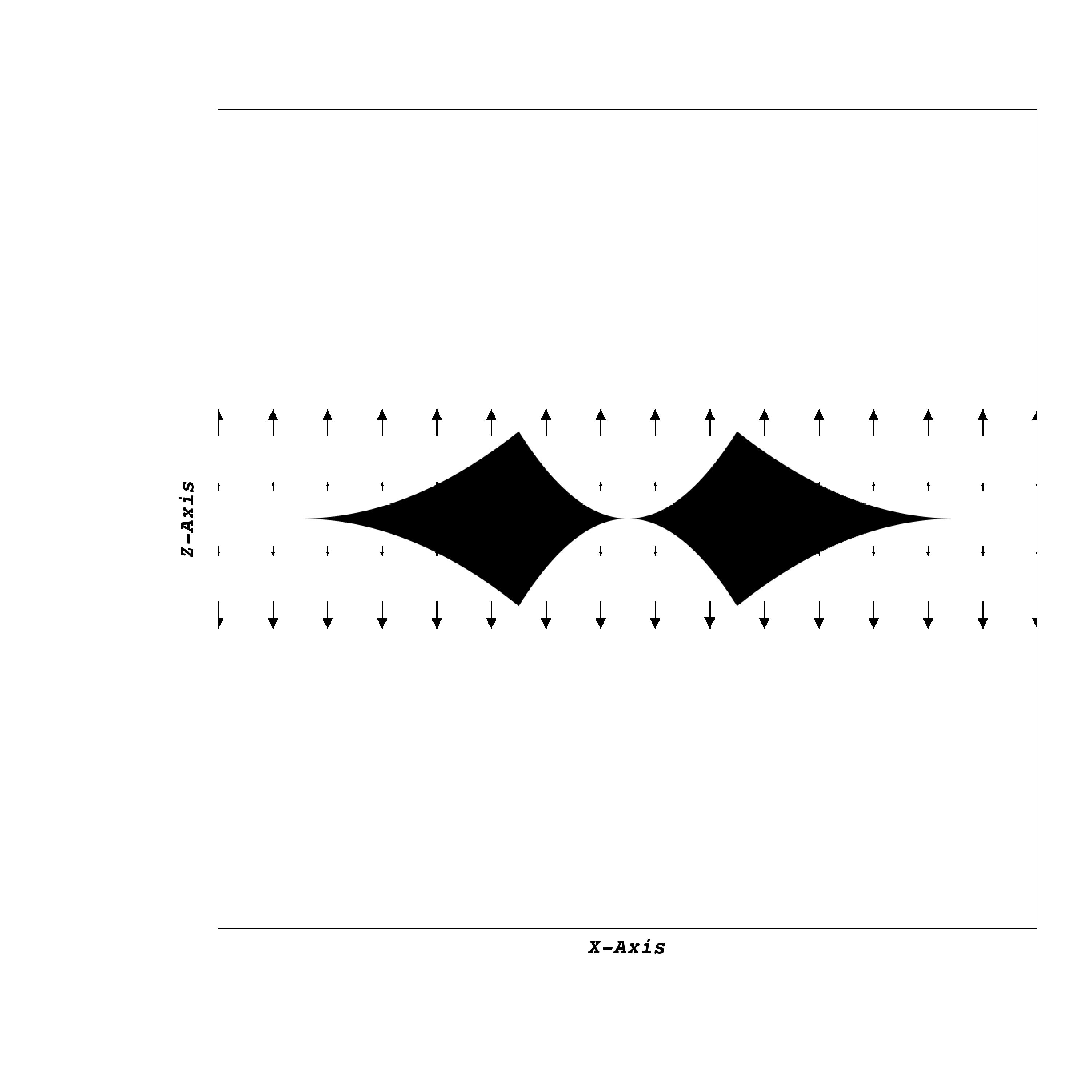}
   \includegraphics[trim=20cm 15cm 0cm 10cm,clip=true,width=6cm]{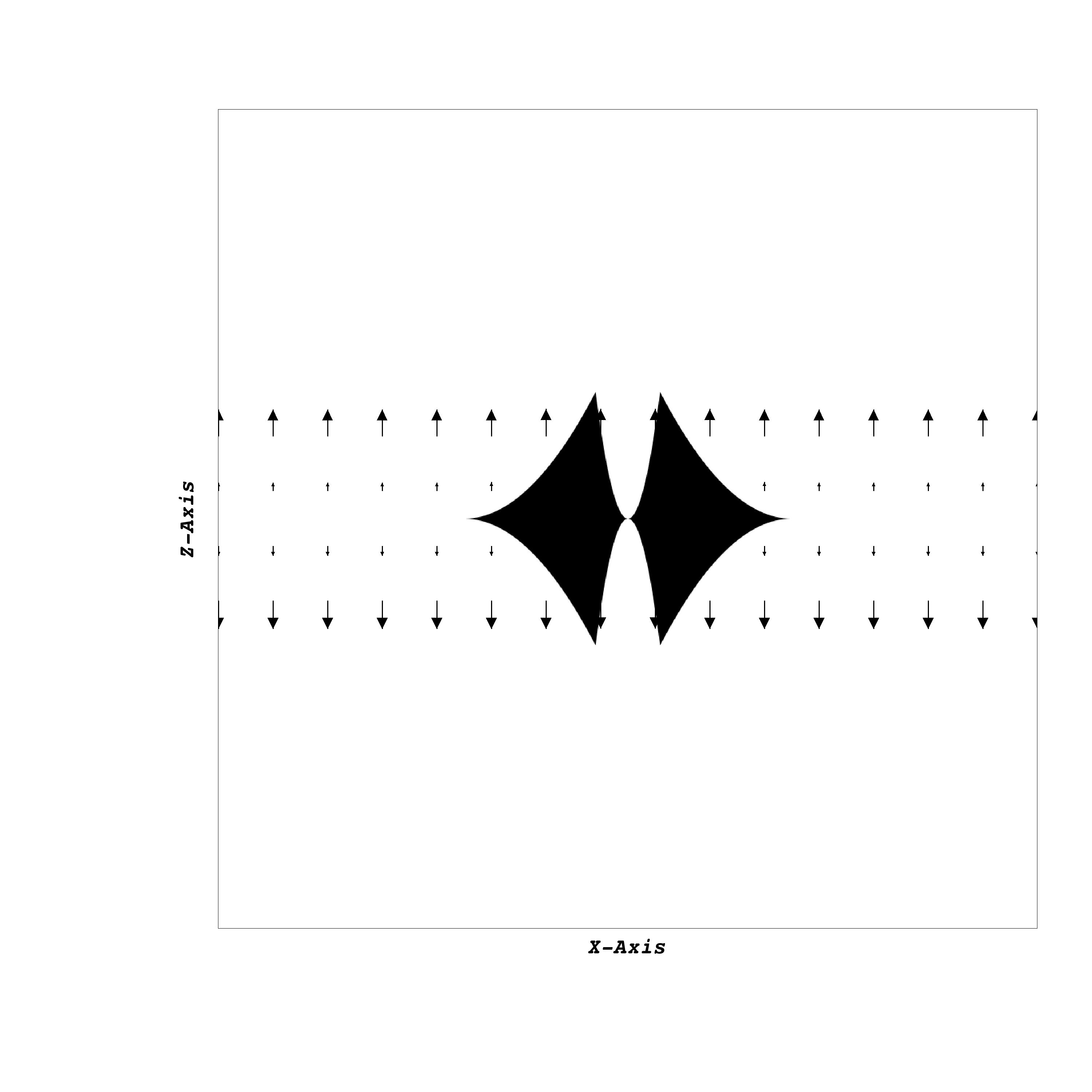}
      \caption{Illustration of the 'Cavity' model with three different parameter sets for the parabolae that define the density distribution. Arrows indicate velocity vectors, contours the density distribution.}
         \label{fig:setup_cavity}
   \end{figure*}

In a later stage, the galactic feedback might have completely swept away or ionized the gas along its path, while in the plane, cold gas is still
accreted. In an extreme scenario, this leads to a quasi-static situation where a small region along the $\pm z$-axis is essentially evacuated. This model is somewhat similar to the shell with holes model, but due to the gaseous torus around the cavity, there is a high probability that photons coming from random angles are scattered so that they are eventually aligned with the $z$-axis. Here, our density profile is described by the intersection of two rotated parabolae, $p_1(x)=a_1(|x|-b_1)^2+c_1$ and $p_2(x)=a_2(|x|-b_2)^2+c_2$ where $x$ is the perpendicular distance from the z-axis. The density at a point at distance $x'$ from the z-axis and at a height $h$ above the plane is non-zero when $h<p_1(x')$ and $h<p_2(x')$.

We define $r_c$ to be the radius of the object at $z=0$ and set $\frac{N_H^{''}}{r_c}= \rho$. We label the column density with two primes because it is not equivalent to the column densities in the models above since the effective column density seen by a photon emitted at an initial angle $\theta_0$ will non-trivially depend on this angle due to the shape of the density distribution (also see Fig. \ref{fig:setup_cavity}) Additionally, we add a linear velocity field 
\begin{equation}
 \vec v = \vec e_z \frac{z}{z_{max}} v_l\,\,,
\end{equation}
which is proportional to the height over the plane and parallel to the $\pm z$-axis. $z_{max}$ is the $z$-coordinate of the 'tip' of the density distribution. This setup is illustrated in Fig. \ref{fig:setup_cavity}.

\begin{table}
\caption{Summary: Model Parameters}             
\label{table:model_params}      
\centering                          
\begin{tabular}{c c}        
\hline\hline                 
Model & Parameter Set \\    
\hline                        
   Shell & $N_H$, $v$, $b$ \\      
   Bipolar & $N^{'}_H$, $b$, $v_{tot}$, $f$, $t$ \\
   Shell with Holes & $N_H$, $v$, $b$, $\Omega$ \\
   Cavity & $N_H^{''}$, $b$, parameters of the parabolae, $v_l$ \\
\hline                                   
\end{tabular}
\end{table}

\section{Results}

\subsection{Flux and Spectra for Individual Models}\label{Sec:Indiv}
\label{sec:individual}
\begin{figure*}
   \centering
   \includegraphics[width=0.8\linewidth]{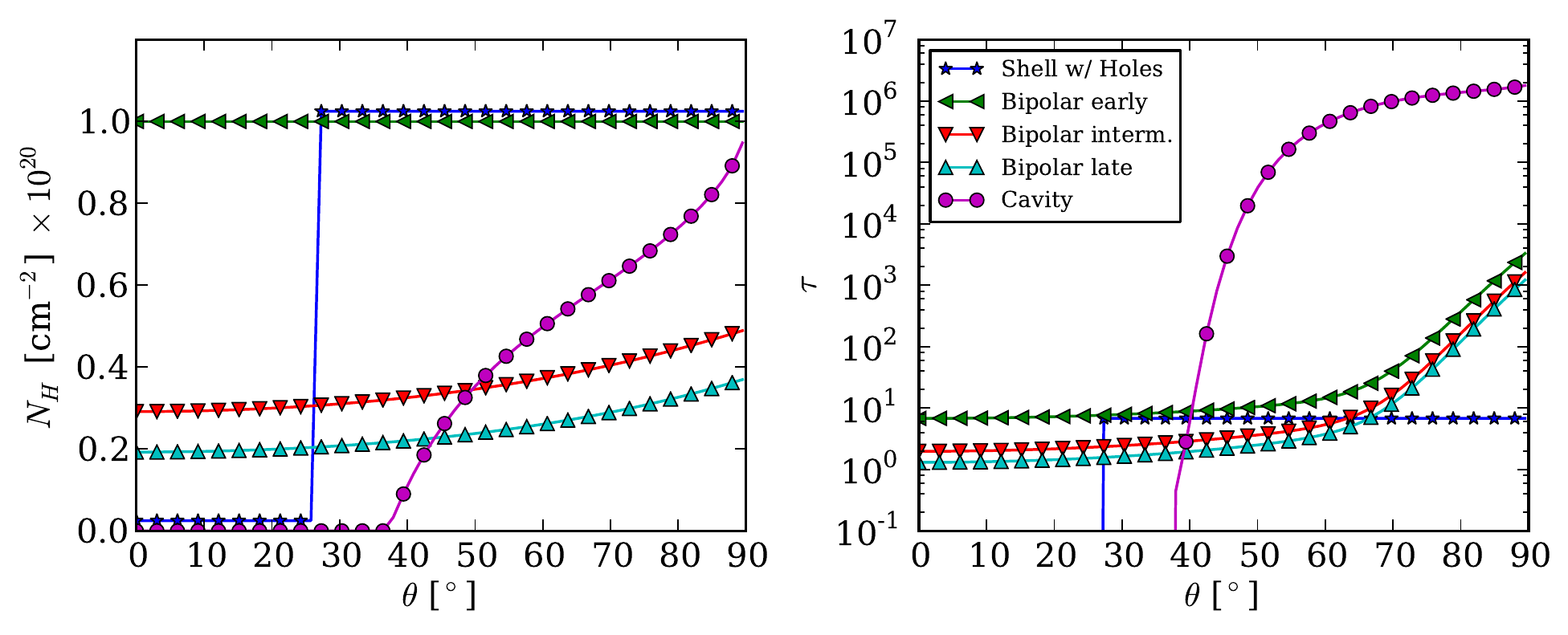}
      \caption{Neutral hydrogen column density (left) and optical depth (right) of photons starting in the center of the setup with $x=0$ for realizations of different models discussed in Sec. \ref{Sec:Indiv} as a function of the angle $\theta$ between the line of sight an the the $z$-axis. Note the log-scaling in the right plot.}
         \label{fig:NH}
   \end{figure*}

In this section, we show results for three individual realizations of our models. If not stated otherwise, we set the Doppler parameter $b$ to 40 km/s and insert the \lya photons with a Gaussian distribution around the line center, where the width $\sigma$ is set to 100 km/s. $10^7$ photons were calculated for every realization to ensure convergence. For the models presented here, we show the column density and initial optical depth (for a photon at line center) as a function of inclination in Fig. \ref{fig:NH}.

\subsubsection{Bipolar}
\begin{figure*}
   \centering
   \includegraphics[width=0.8\linewidth]{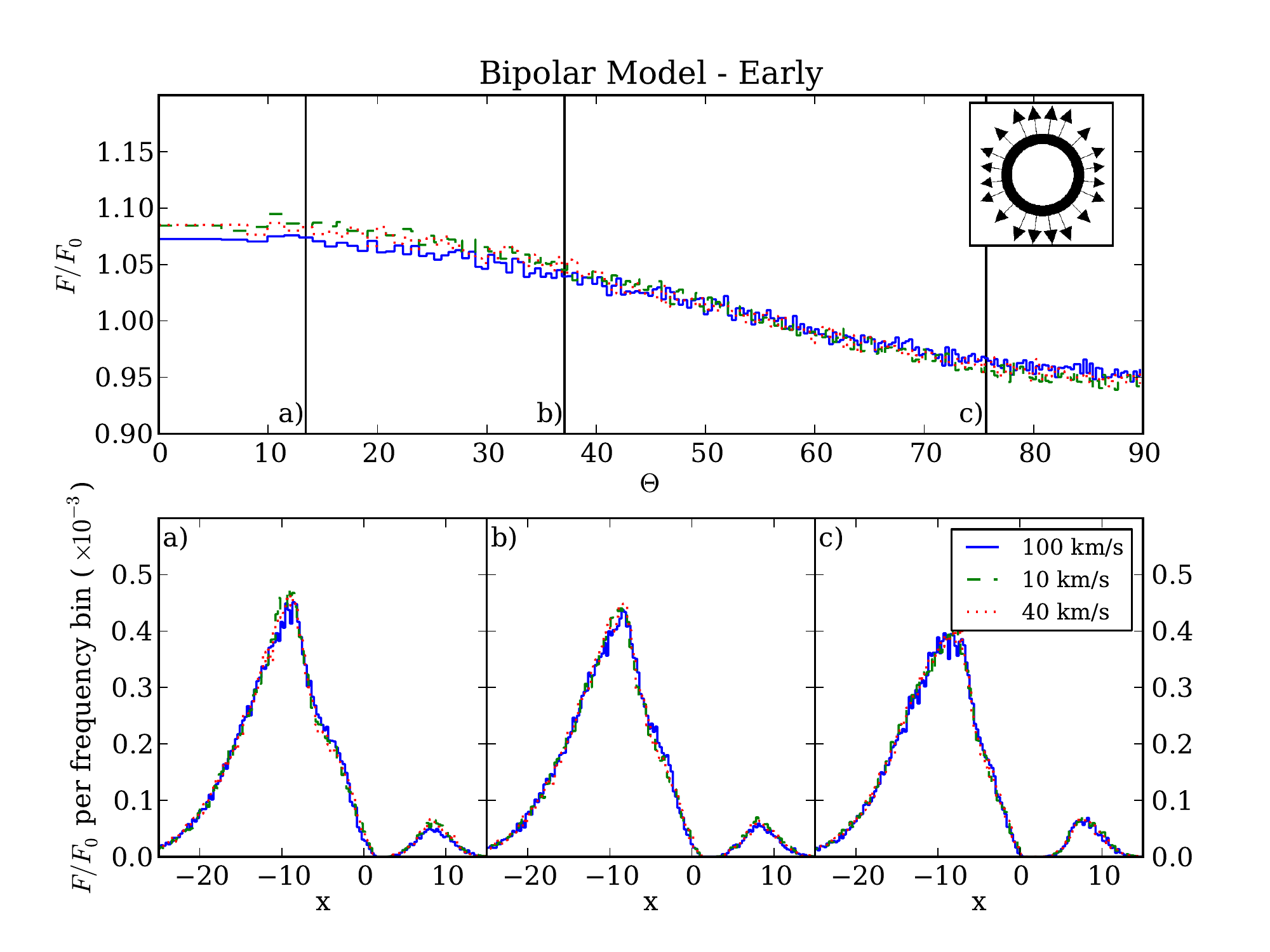}
      \caption{Top: Flux as a function of angle of observation for the shell with holes model. Units are chosen in a way so that a value of 1 would correspond to the isotropic case. Bottom: Spectrum evaluated at three different angular bins (indicated by vertical lines above). Results are shown for input spectra with a width of 100 (blue,solid), 10 (green, dashed) and 40 (red,  dotted) km/s. Spectra are normalized to the total flux in the angular bin. Model parameters are $N^{'}_H=10^{20}$ cm$^{-2}$, $v_{tot}=200$ km/s, $f_b=0.5$, $t=0$}
         \label{fig:bi_early_dist}
   \centering
   \includegraphics[width=0.8\linewidth]{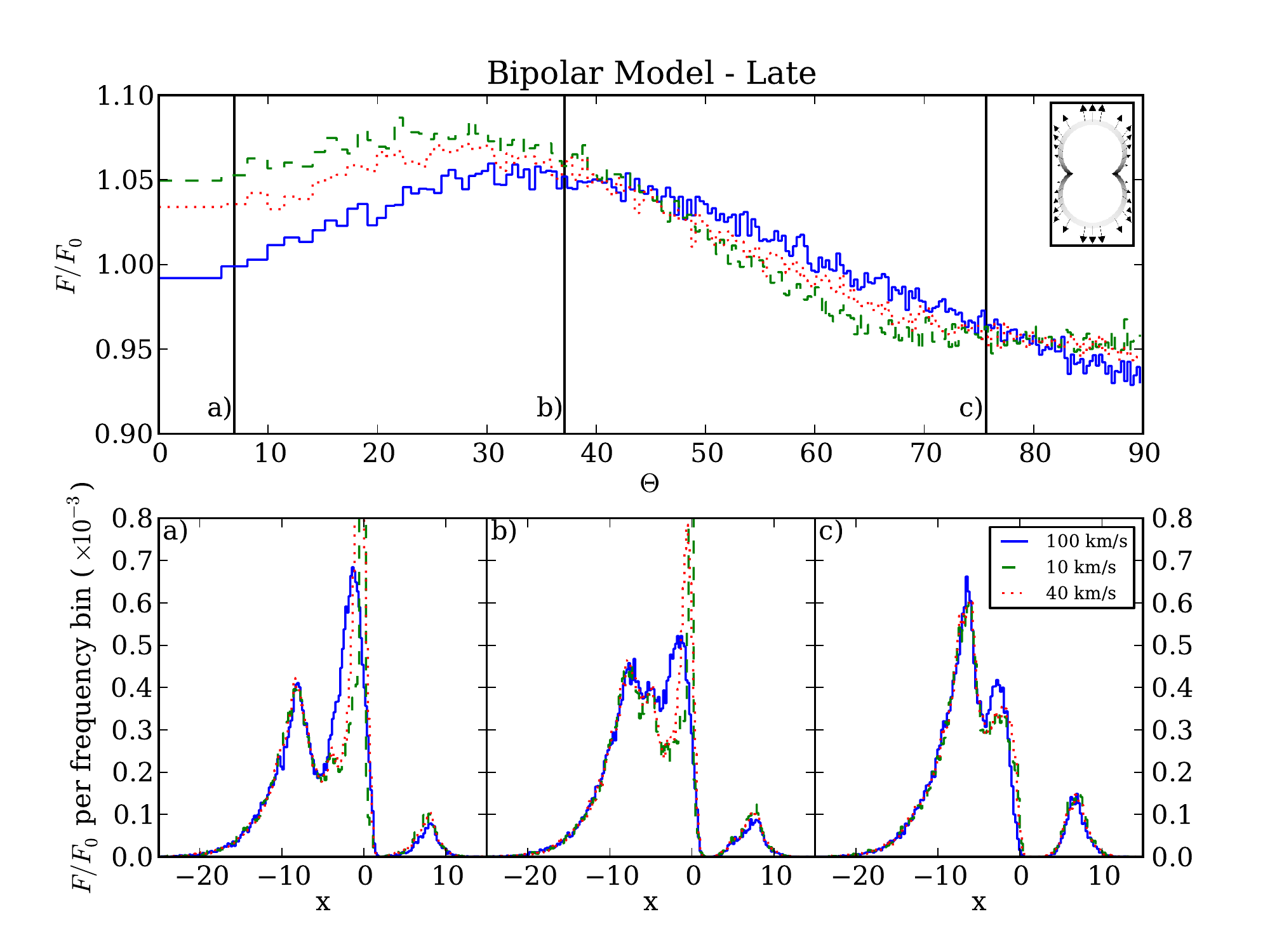}
      \caption{Same as Fig. \ref{fig:bi_early_dist}, but for the late bipolar model. Model parameters are $N^{'}_H=10^{20}$ cm$^{-2}$, $v_{tot}=200$ km/s, $f_b=0.5$, $t=0.15$}
         \label{fig:bi_late_dist}
   \end{figure*}

In the upper panel of Fig. \ref{fig:bi_early_dist}, we show the flux as a function of viewing angle for a particular case of the bipolar model ($v_{tot}=200$ km/s, $f_b=0.5$, $N^{'}_H=10^{20}$ cm$^{-2}$, $t=0$). 
Values are given in units of the corresponding isotropic distribution among the observers. Let $F=n/N$ be the relative flux an observer detects, where $n$ is the number of photons received by the observer and $N$ the total number of photons simulated. Further, let $F_0 = N/n_{bins}$ be the relative flux assuming isotropy with $n_{bins}$ the number angular bins. The flux in these bins can be seen as the flux an observer at a specific inclination would obtain. $F/F_0$ measures the relative flux relative to the isotropic case. We stress that ssuming the object to be dust-free, this quantity is proportional to the EW of the observed \lya line.

Since for the setup in Fig. \ref{fig:bi_early_dist} the time parameter is zero, the density field is equivalent to the isotropic shell model, but the velocity field exhibits an anisotropy, lowering the optical depth along the $z$-axis by a factor of $10^3$ with respect to lines of sight within the $xy$-plane. As can be seen in the upper panel of Fig. \ref{fig:bi_early_dist} the flux variation is small, about 15\% more photons escape at $\theta\approx 0^\circ$ than at $\theta\approx 90^\circ$. In the bottom panels of Fig. \ref{fig:bi_early_dist}, we show spectra obtained at three different observations angles. Horizontal lines in the top panel show at which angles the spectra where evaluated. For these parameters, the shape of the spectrum does not change significantly with observation angle. The three different lines Fig. \ref{fig:bi_early_dist} show the results for different input spectra with a width of 100/40/10 km/s (solid/dotted/dashed lines). As expected, a narrow input spectrum leads to a higher and 
narrower peak for small observation angles. The flux does not change significantly with respect to the width of the Gaussian.

In Fig. \ref{fig:bi_late_dist}, we show the same quantities for a more evolved setup with $t=0.15$ which will we will refer to as the 'late bipolar' case. Here, the flux distribution shows a counterintuitive trend; while flux is suppressed for large observations angles, it is also suppressed for very small angles, leading to a maximum at around $\theta=38^\circ$. Detailed investigation shows that this is a geometrical effect from the deformation of the shell. Most parts of the surface of the bipolar shell point towards the $xy$-plane. Photons escaping the shell in these areas will most likely not escape with a small angle $\theta$. Photons that do escape with small angles most probably escape from the poles. This leads to enhanced flux at intermediate angles, because photons escaping at these angles can statistically escape at every point of the surface, while those at small/large angles are confined to certain areas. This could also be described as a 'projection effect'. Our explanation is further supported 
by the fact that the effect becomes stronger for a broader input spectrum. The blue part of the Gaussian is subject to scatterings even for photons traversing the shell along the $z$-axis. The broader the spectrum, the more blue photons are scattered at the poles, redistributing their orientation and further enhancing measured flux at intermediate angles.

The spectra shown in the bottom panels in Fig. \ref{fig:bi_late_dist} again show a peak at $x=0$ for observers looking down the $z$-axis (bottom left panel). For an input spectrum with a width of 100 km/s, it is slightly shifted. This can be attributed to the suppression of the blue part of the input spectrum due to scatterings in the outflowing material. For smaller widths, the peak is clearly centered at $x=0$ again. For observers around $\theta=40^\circ$, the amplitude of the peak is lower. If we go to larger angles, the central peak vanishes and the red peak visible in all of the three spectra becomes dominant.

\subsubsection{Shell with Holes}
\begin{figure*}
   \centering
   \includegraphics[width=0.8\linewidth]{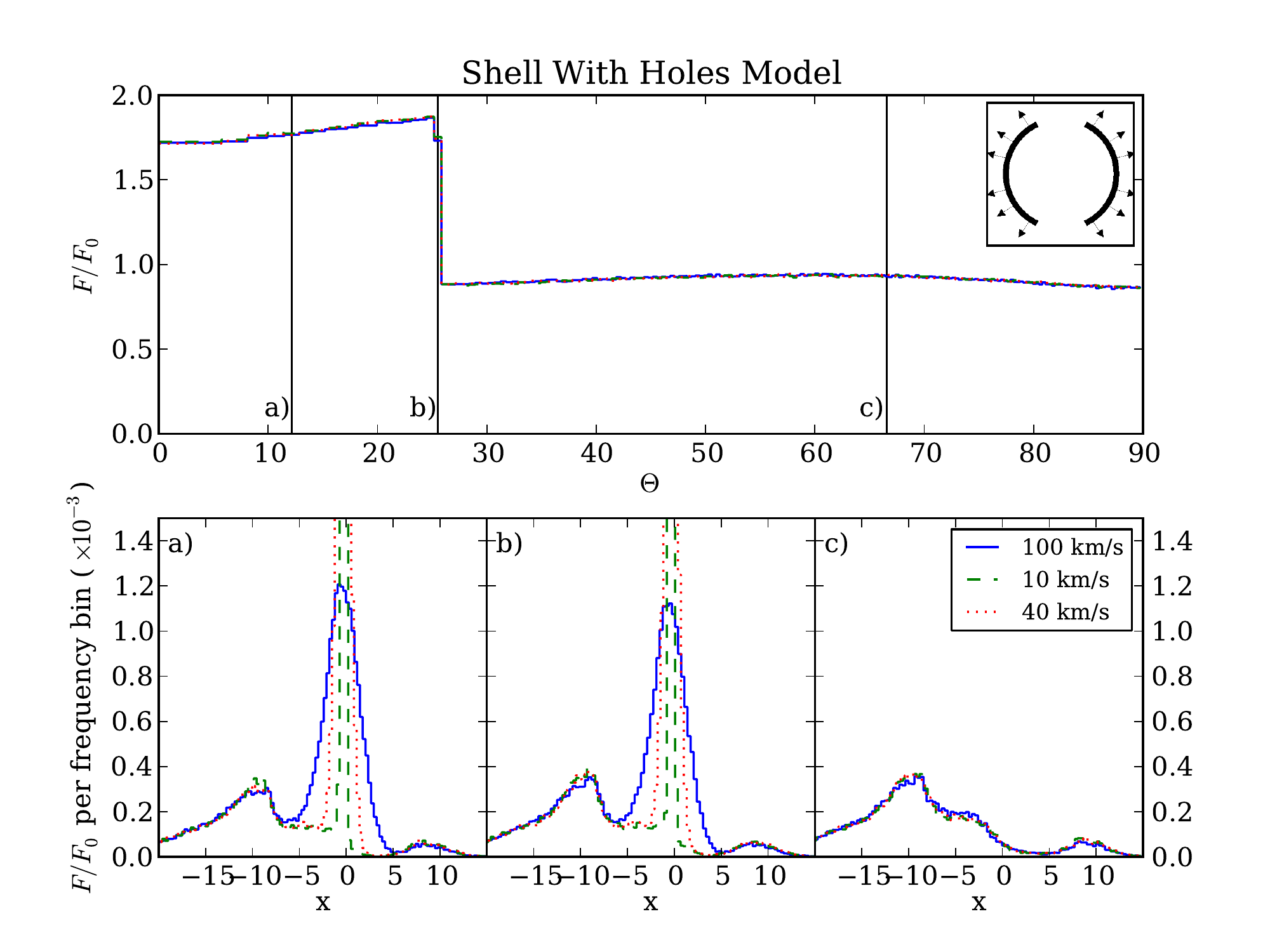}
      \caption{Same as Fig. \ref{fig:bi_early_dist}, but for the early bipolar model. Model parameters are $v=200$ km/s, $N_H=10^{20}$ and $\Omega=10\%$, meaning that 10\% of the shell is subtended by the holes.}
         \label{fig:holes_dist}
   \end{figure*}

The dependence of flux and spectra for one particular case ($v=200$ km/s, $N_H=10^{20}$ cm$^{-2}$, $\Omega=10\%$) of the shell with holes model is shown in the upper panel of Fig. \ref{fig:holes_dist}. 

As can be seen in Fig. \ref{fig:holes_dist}, the observed flux decreases abruptly at an angle of $\approx 25^\circ$, which reflects the angular size of the holes in the shell. In the lower panels of Fig. \ref{fig:holes_dist}, we show spectra of the photons in three different angular bins. Their positions are indicated by vertical lines in the upper panel. The bottom left spectrum shows the spectrum for an observer that directly looks ``down the hole''. As expected, one can clearly identify the unprocessed, Gaussian input centered around $x=0$. The spectrum in the bottom center shows a spectrum directly after the transition; the transition itself occurs so fast that we do not see it occuring in one of the bins. 
 The bottom right spectrum shows an observer at an angle of $\approx 67^\circ$. Here, the peak at line center is clearly gone. Instead, the spectrum more resembles the original shell case. We note that observers looking ``down the hole''  see a \lya line that is not shifted. In this particular setup, observing face-on will boost the flux and therefore the EW by a factor of roughly $1.7$ (assuming no scattering of the continuum emission here). It is interesting that apart from the sharp transition at the hole's half opening angle, the spectrum does not evolve significantly. This can be seen by comparing the center bottom and right bottom spectrum. We stress that both the flux and the spectrum change very rapidly and are nearly constant before and after the transition. Beyond the transition, the spectrum resembles the normal shell case. Before the transition, it consists of two components, the initial unscattered spectrum and another contribution from a shell-like spectrum. The latter contribution 
comes from photons that merely scatter into the line of 
sight. 
This implies that for decreasing hole radius the spectrum asymptotes toward the original shell case.


\subsubsection{Cavity Model}
\begin{figure*}
   \centering
   \includegraphics[width=0.8\linewidth]{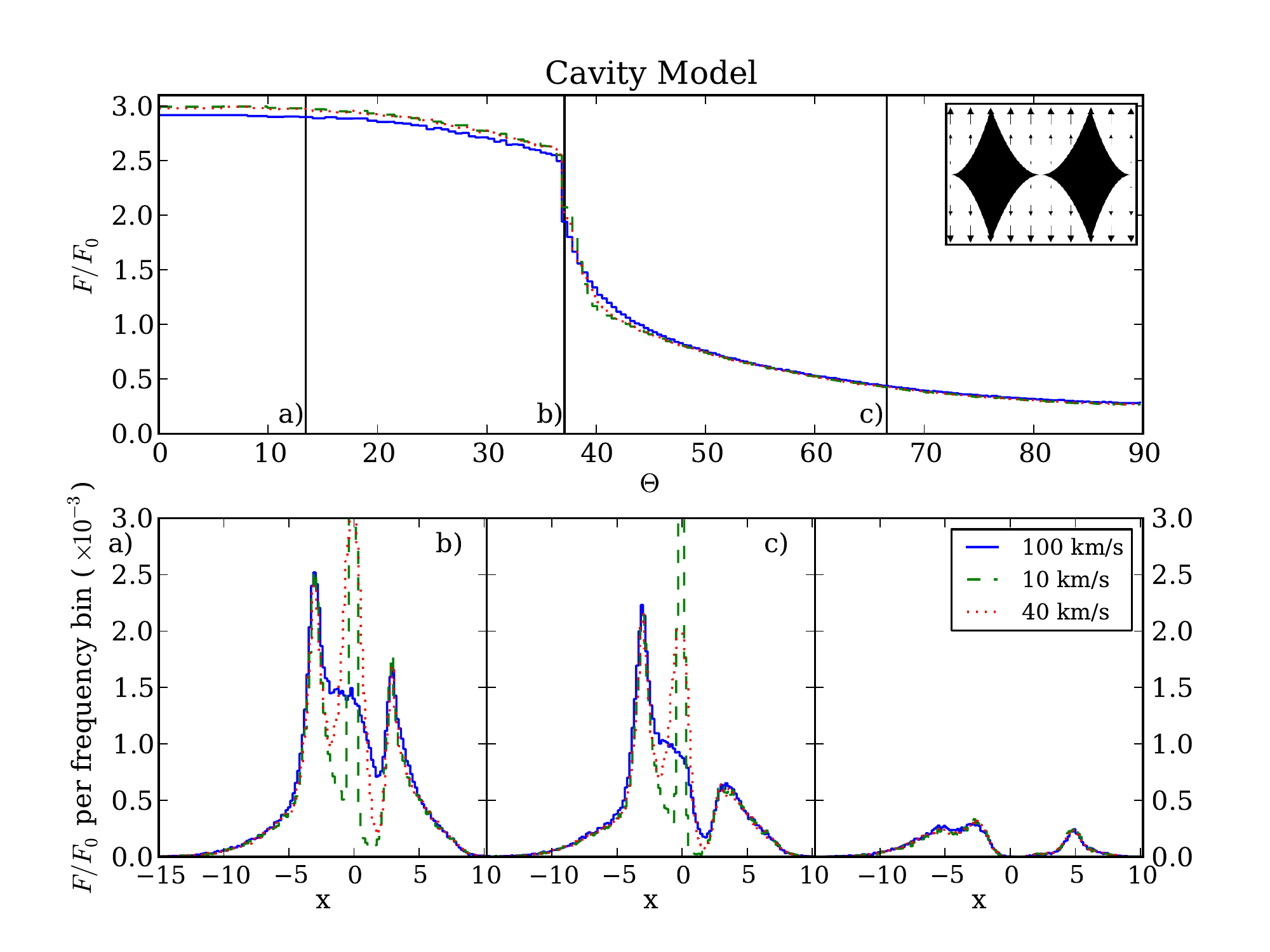}
      \caption{Same as Fig. \ref{fig:bi_early_dist}, but for the cavity model. Model parameters are $N^{''}_H=10^{20}$ cm$^{-2}$, $v_l=200$ km/s.}
         \label{fig:beaming_dist}
   \end{figure*}

In the upper panel of Fig. \ref{fig:beaming_dist} the flux distribution for one realization of the cavity model is shown. The model parameters here are $N^{''}_H=10^{20}$ cm$^{-2}$, $v_l=200$ km/s, we use the configuration illustrated in the left plot of Fig. \ref{fig:setup_cavity}. Similar to the shell with holes model, we get a sharp transition at some ``effective opening angle''. Here, flux is suppressed up to a factor of 6 for observers looking edge-on, which as discussed above would also correspond to a boost/suppression of the EW by a factor of 3 with respect to what one expects in an isotropic environment.
When looking at the three spectra shown at the bottom of Fig. \ref{fig:beaming_dist}, one notices that in contrast to the shell with  holes case, the spectral properties do not change abruptly. The amplitude of the peaks is reduced with viewing angle, and the overall amplitude shrinks, but the input spectrum is not visible anymore at least for the fiducial case with an width of the input spectrum of 100 km/s (solid). The reason for this different behaviour is that while observers near face-on position still see a significant fraction of photons that directly traversed the cavity, the amount of photons that merely scattered into the line of sight is much bigger. Looking closely at the left and center spectra, one can see the right shoulder 
of the Gaussian input spectrum. This modified Gaussian is gone far away form the transition, which leads to negligible flux around $x=0$, as can be seen in the right bottom plot of Fig. \ref{fig:beaming_dist}. For narrower input line, we recover the input spectrum again. We conclude that while we get a large flux suppression for this setup, observers could only distinguish from the spectra alone whether or not they are looking down the hole or not if the initial line is much narrower than 100 km/s. The maximum amplitude of a broad Gaussian is not bright enough to be distinguished from the rest of the spectrum.

We note that in contrast to the late bipolar model, we do not see a suppression of the flux at small angles due to geometrical effects although the density field in this case is clearly deformed as well. We attribute this to the fact that the differences in optical depth among different lines of sight are much larger than in the discussed bipolar setup. In contrast to the shell with holes model, in this case the transition in the spectral shape and the flux is not as rapid, but more smoothly. This is can be explained by considering the smooth transition in column density and optical depth as shown in Fig. \ref{fig:NH}, as well as by the fact that the escape mechanism for the cavity model is different (see below).

\subsection{The Influence of Dust}
\label{sec:dust}
\begin{figure*}
   \centering
   \includegraphics[width=0.8\linewidth]{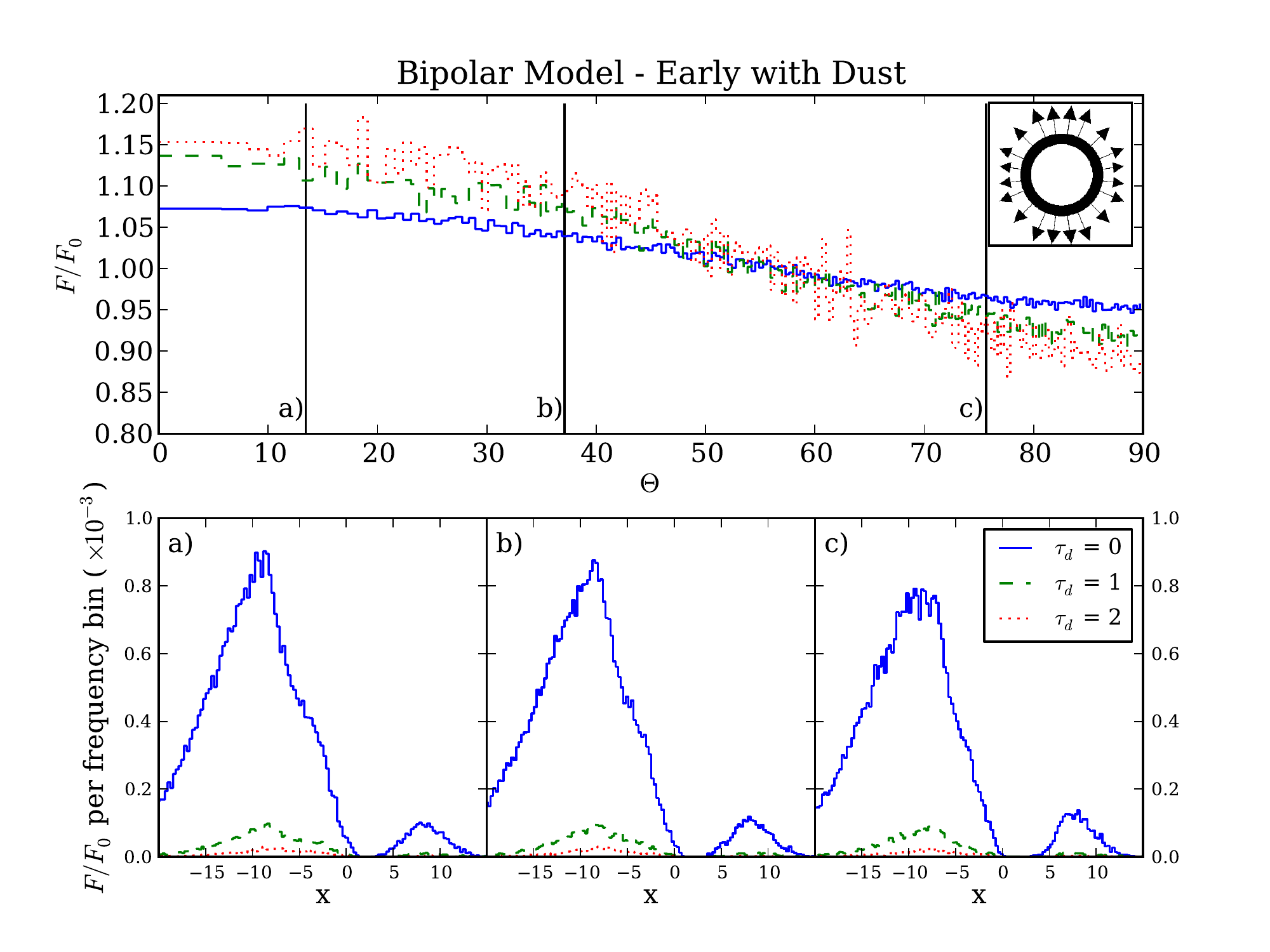}
      \caption{Same as Fig. \ref{fig:bi_early_dist}, but for three different optical depths in dust: $\tau_D=$0 (blue, solid), 1(green, dashed), 2(red, dotted). Model parameters are $N^{'}_H=10^{20}$ cm$^{-2}$, $v_{tot}=200$ km/s, $f=0.5$, $t=0$.}
         \label{fig:bi_early_dist_dust}
%
   \centering
   \includegraphics[width=0.8\linewidth]{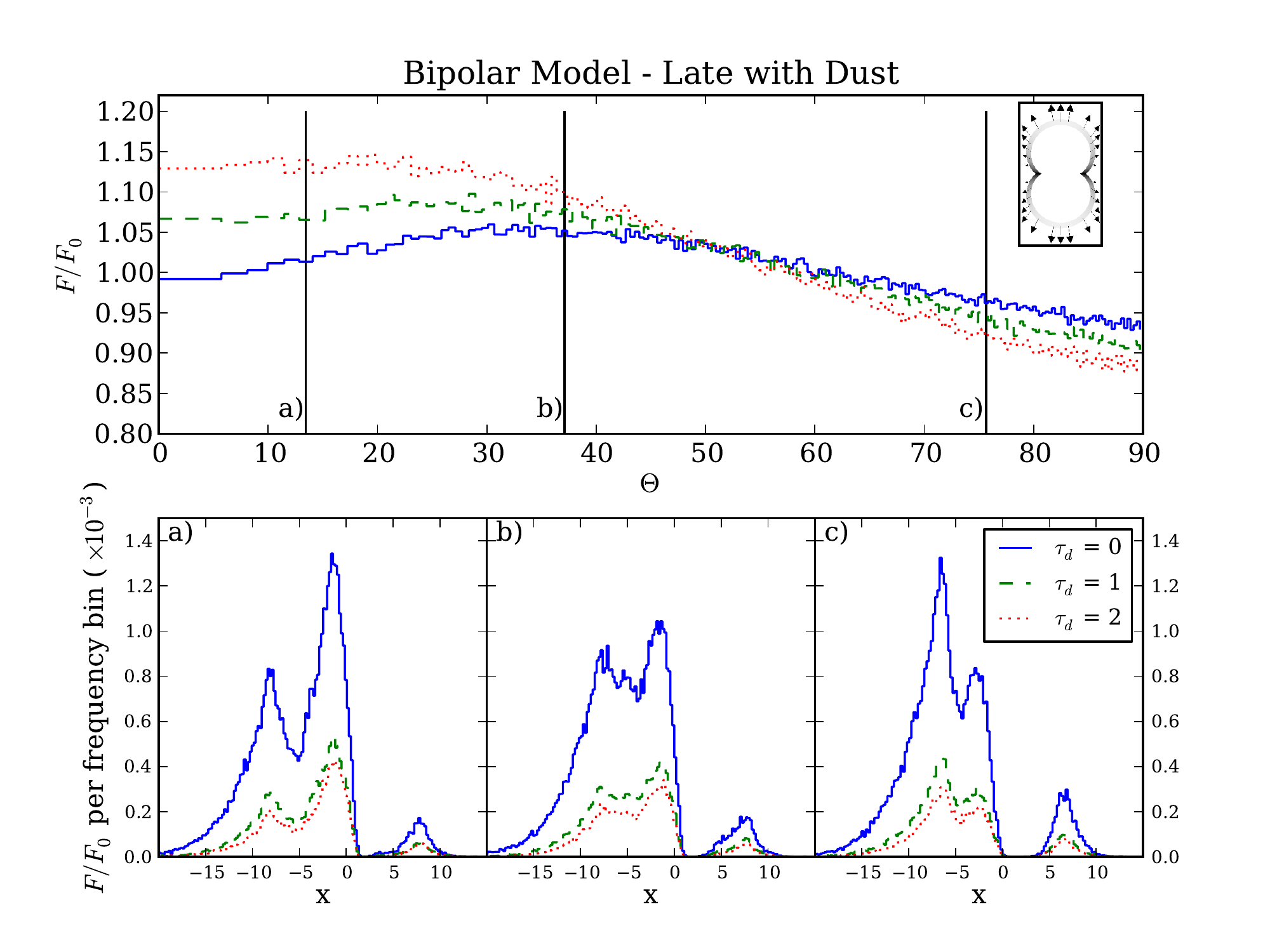}
      \caption{Same as Fig. \ref{fig:bi_early_dist_dust}, but for the late bipolar model. Model parameters are $N^{'}_H=10^{20}$ cm$^{-2}$, $v_{tot}=200$ km/s, $f=0.5$, $t=0.15$}
         \label{fig:bi_late_dist_dust}
   \end{figure*}

\begin{figure*}
   \centering
   \includegraphics[width=0.8\linewidth]{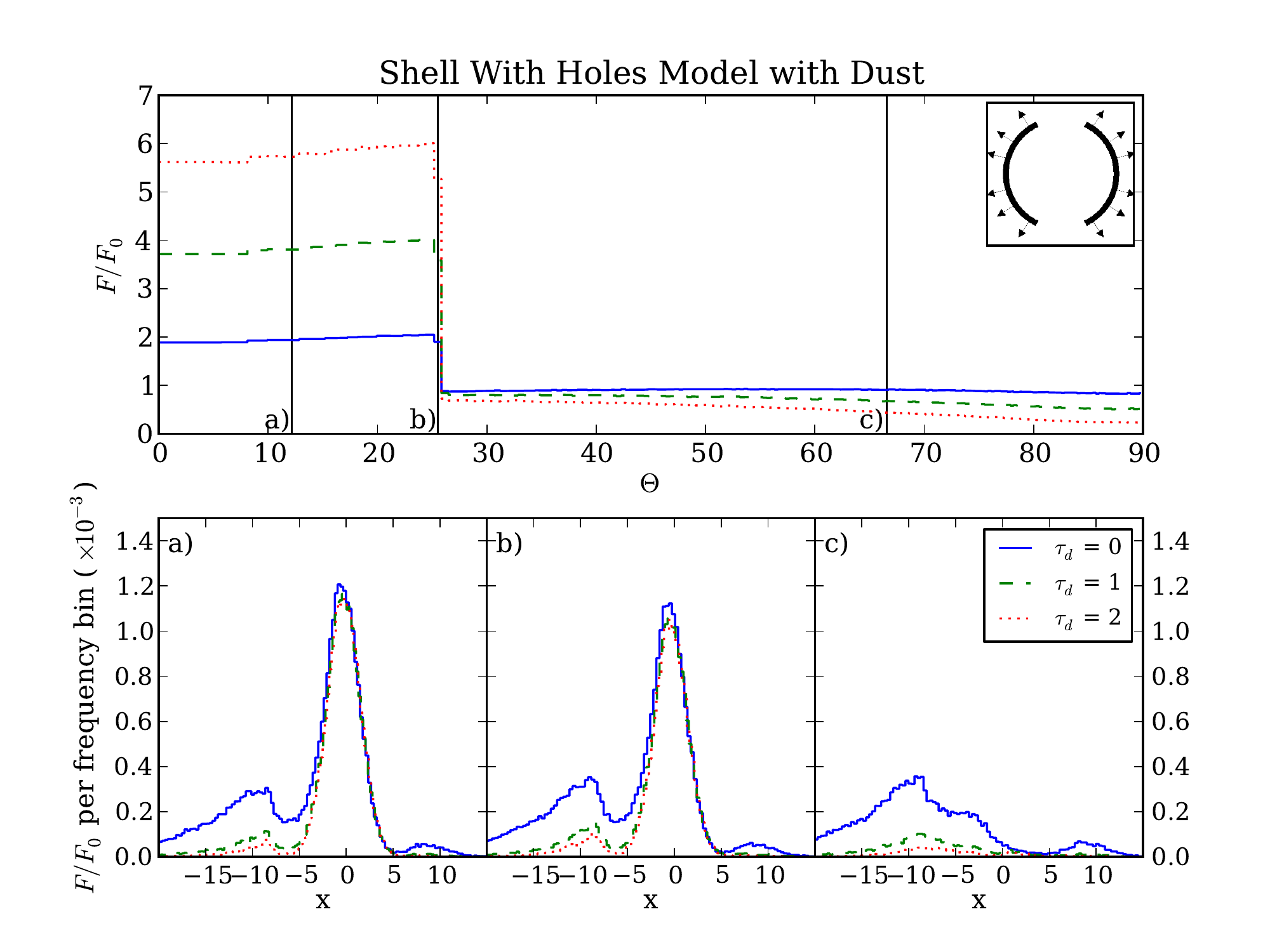}
      \caption{Same as Fig. \ref{fig:bi_early_dist_dust}, but for the shell with holes model. Model parameters are $v=200$ km/s, $N_H=10^{20}$ and $\Omega=10\%$.}
         \label{fig:holes_dist_dust}
   \end{figure*}

\begin{figure*}
\centering
\includegraphics[width=0.8\linewidth]{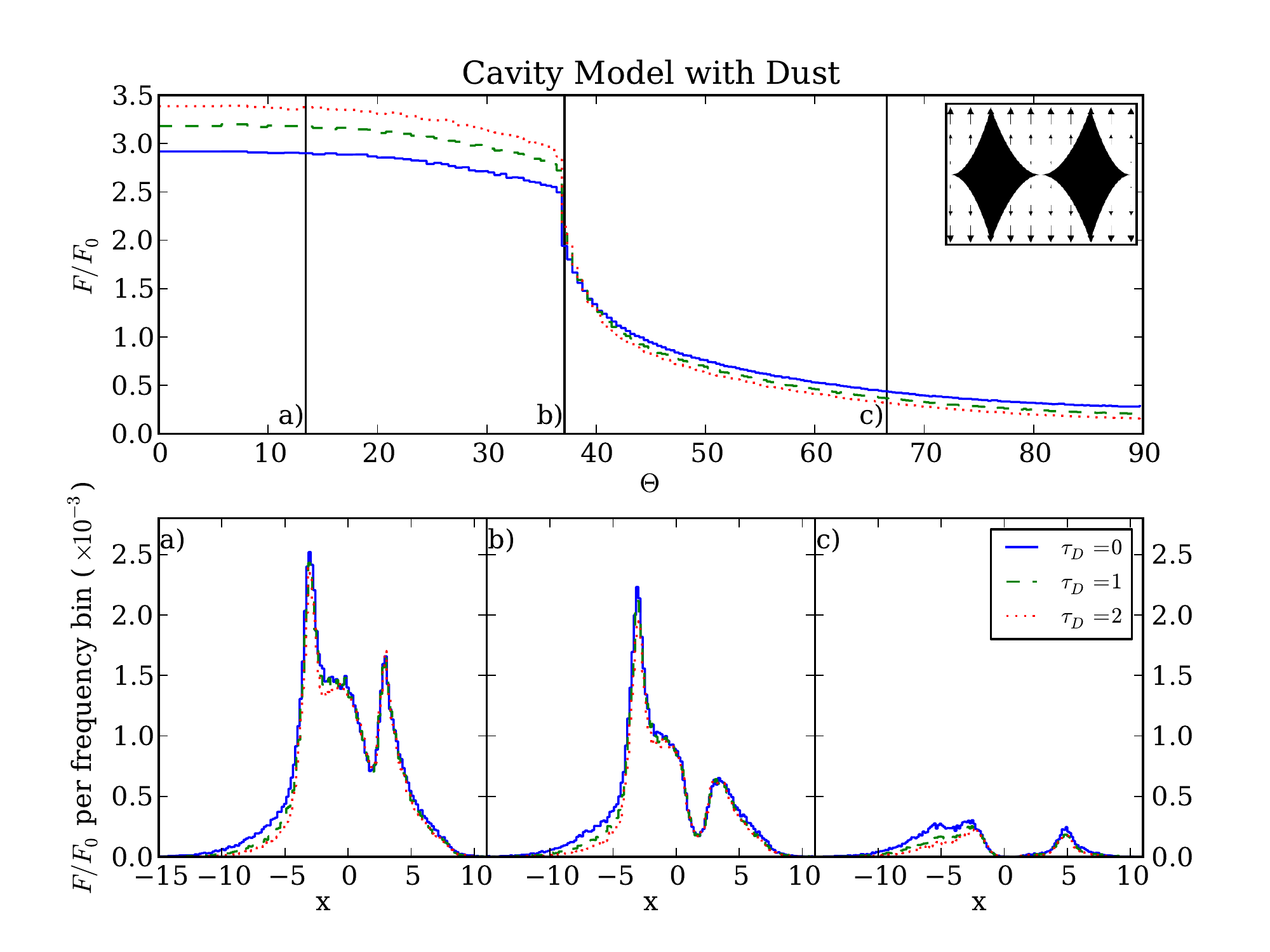}
  \caption{Same as Fig. \ref{fig:holes_dist_dust}, but for the cavity model. odel parameters are $N^{''}_H=10^{20}$ cm$^{-2}$, $v_l=200$ km/s.}
      \label{fig:beaming_dist_dust}
\end{figure*}

While not being the main focus of this paper, the influence of dust on the aforementioned phenomena is also of interest. Therefore, we reran the individual cases presented in the previous sections with varying dust content. We introduced dust to be proportional to hydrogen content, following the implementation given in \cite{Verhamme06} with an albedo of 0.5 and an isotropic phase function in case of scattering on dust. We quantify the dust content in terms of the additional optical depth $\tau_D$. Similar to the hydrogen column density, $\tau_D$ is the depth through the shell for the shell with holes model, the optical depth at $t=0$ for the bipolar models and the optical depth in the plane for the cavity model. The results are shown in Fig. \ref{fig:bi_early_dist_dust}, \ref{fig:bi_late_dist_dust}, \ref{fig:holes_dist_dust}, and \ref{fig:beaming_dist_dust}. All runs were perfomed using an input spectrum of width 100 km/s, and assuming $b= 40$ km/s.

As expected, the dependence of the flux distribution on inclination in general becomes larger for higher dust content. For the shell with holes case in Fig. \ref{fig:holes_dist_dust}, we see a boost in the flux anisotropy by a factor of 3 with respect to the dust-free case: photons not escaping through the holes are subject to destruction by dust, so relatively the flux through the holes increases. The same accounts for both the bipolar cases in Fig. \ref{fig:bi_early_dist_dust} and \ref{fig:bi_late_dist_dust}, although the effect is less strong. Interestingly, dust does not strongly influence the cavity model as shown in Fig. \ref{fig:beaming_dist_dust} if one compares to the shell with holes case. Surprisingly, the $\tau_D = 2$ just increases the flux boost by a factor $\approx 20\%$. Analysis of the spatial distribution with regard to the point of last scattering of escaping photons show that most of the photons bounce of the inner trough of the density distribution without penetrating deeply, so most 
photons do not 
experience 
the dust opacity. This escape mechanism also partly explains the smooth transition of the spectra/flux distribution mentioned before.

\subsection{Parameter Study: Peak Position versus Observation Angle}
\label{sec:pstudy}
\begin{figure}
   \centering
   \includegraphics[width=\linewidth]{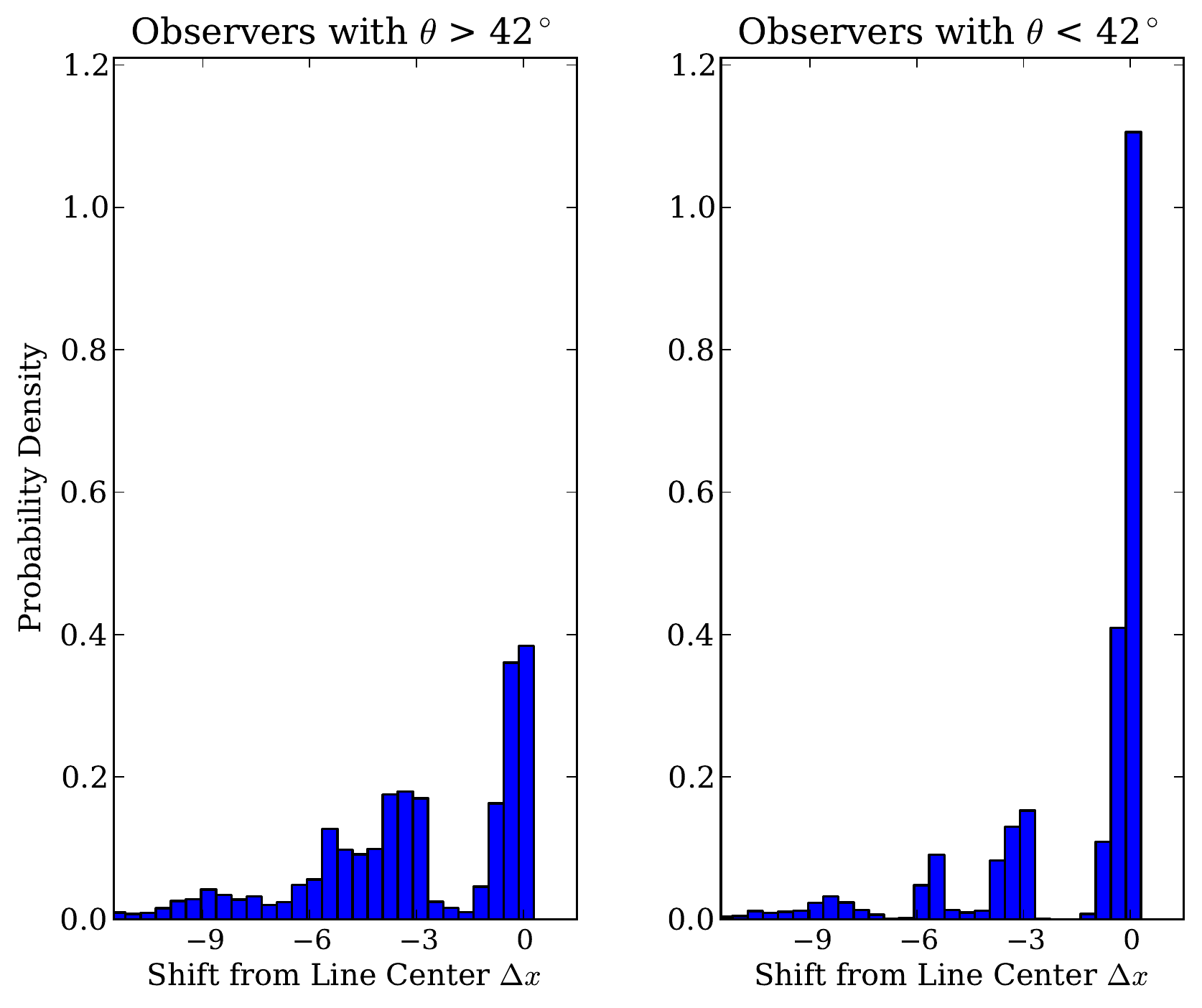}
      \caption{Distribution of the spectral peak for our parameter study, see text for details. Left: Distribution of peak position for observers with $\theta > 42^\circ$. Right: Distribution for observers with $\theta < 42^\circ$.}
         \label{fig:parameter_study_dist}
   \end{figure}

To examine the correlation of a peak at the systemic velocity (at $x=0$) with low inclination, we ran a suite of models with different parameters. We set the width of the input spectrum to 40 km/s here. The parameter range for the three different model families is shown in table \ref{table:parameter_space}. In total, we ran 272 models, all without dust. Since certain parameter combinations lead to weak anisotropies in optical depth and column density, we restricted our analysis to those for which either the flux changed by at least 10\% with respect to the observation angle and/or for which the highest peak in the spectrum moves at least 3 units in frequency depending on observation angle. We stress that by these conditions, we do not select only those models that already show the aforementioned correlation, but all those models that show anisotropic spectra and/or fluxes in general. We end up with a total of 179 models. In Fig. \ref{fig:parameter_study_dist}, we show the distribution of the peak position 
for those models. In the left panel, the distribution is 
shown for observers at angular positions $\theta > 42^\circ$, while in the right panel, we show the distribution for observers within $\theta < 42^\circ$. For the latter, there is a strong peak at $x\approx0$. While for observers above $42^\circ$ there is a peak at the systemic velocity as well, it is much smaller than for observers below $42^\circ$. Within the scanned region of parameter space, we therefore find a correlation between peak shift and inclination. For low inclination, the probability to see a peak at $x\approx0$ is enhanced by a factor of a few. This trend is consistent in all our three model families, but stronger for the shell with holes and cavity models. 
This suggests that a Ly$\alpha$ spectrum with a peak at $\Delta x \sim 0$ is more likely (than random) to be associated with a sightline passing through a hole or cavity. We stress that due to the finite spectral resolution of observations, the peaks can be washed out. In this case, we don't expect a distinct peak at systemic velocity, but a flux excess near the systemic velocity.

\begin{table*}
\caption{Summary: Explored Parameters}             
\label{table:parameter_space}      
\centering                          
\begin{tabular}{c c}        
\hline\hline                 
Model & Parameter Set \\    
\hline                        
   Bipolar & $N^{'}_H$=[10$^{19}$, 10$^{20}$, 10$^{21}$ cm$^{-2}]$, $v_{tot}$= [100, 200, 300 km/s], $f$=[0.25, 0.5, 0.75, 1.0], $t$=[0,0.05,0.1,0.2] \\
   Shell with Holes & $N_H$=[10$^{19}$, 10$^{20}$, 10$^{21}$ cm$^{-2}]$, $v=$ [100, 200, 300 km/s], $\Omega=$[5\%, 10\%, 25\%, 30\%] \\
   Cavity & $N_H^{''}$=[10$^{19}$, 10$^{20}$, 10$^{21}$ cm$^{-2}]$, $v_l$=[0, 100, 200, 300 km/s] \\
\hline                                   
\end{tabular}
\end{table*}

\section{Discussion \& Conclusions}
We performed Monte-Carlo radiative transfer simulations for the \lya emission in three different families of models that are all inspired by galactic winds and expanding \ion{H}{I} shells. In contrast to the shell models \citep[e.g ][]{Verhamme06}, we have added anisotropic velocity and density components. While the bipolar and shell with holes models are natural extensions to the isotropic shell models, the cavity models are designed to be more extreme with large anisotropies (and reminiscent of a wind breaking out of a galactic disk).

We find that the anisotropies in our models lead to an anisotropic escape of photons which is visible in the spectral properties as well as in the flux as obtained for observers at different angular positions. 
We generally find that the viewing angle-dependence of the predicted spectrum is quite complicated, with the flux in different peaks in the spectrum changing in some cases (see e.g. Fig.~10), but not in others (see e.g. Fig.~9).  Additionally, we find that the inclination effect due to anisotropic density and velocity fields can be partly counteracted by geometrical effects, e.g. for continously deformed shells like in the bipolar late-type cases.

In spite of these complications: there are several results to take away from our analysis:

\begin{itemize}
\item In terms of the EW, we find boosts of up to $\sim$10\% for the bipolar model, boosts of $\sim$2 for the shell with holes model and a factor of $\sim$3 for the cavity model. The inclusion of dust enhances the boost for the bipolar [shell with holes] model to $\sim$15\% [a factor of $\sim 5.5$]. For the cavity model, the inclusion of dust increases the boost to about 3.5.

\item For all of the discussed models, we find a range of parameters for which there is a clear correlation between an observed peak at a velocity shift of zero and an inclination close to zero. This is intuitive, since when looking down a cavity, we expect to see the unaltered spectrum of the emitting region, not strongly affected by scatterings in intervening gas. This can be of practical importance when searching for ionizing flux leaking out of high-redshift galaxies, because the existence of such cavities will also allow ionizing photons to escape the galaxy. Our models suggest that there is an increased probability to find ionizing photons leaking from LAEs for which the \lya spectrum exhibitis a peak at 
systemic velocity. Due to the finite spectral resolution of observations, this peak might be washed out and show as flux excess near the systemic velocity. It is interesting to note that this peak corresponds to the third peak in some of our bipolar and shells-with-holes models.  That this result holds up even when the total change in flux is small; for the bipolar models, we see the clear signature of such a peak even though the total flux variation is as small as 10\%.

Having such an observational diagnostic for sightlines going through low \ion{H}{I} column densities is important, as it may provide a new way to search for so-called `Lyman continuum leakers': the search for these galaxies has been highly inefficient as it requires deep spectroscopic observations of galaxies, for most of which no ionizing flux is detected. Thus far, only a handful of these galaxies have been discovered, and consequently the observational constraints on the escape of ionizing photons from star forming galaxies are weak at best. Since this escape fraction plays a key role in the reionization process, it is important to find observational signatures that may help us constrain this quantity (see also \citet{Nakajima13} and\citet{Jaskot2013} who show that galaxies with high OIII/OII line ratios are more likely to be sources that leak ionizing photons).

\item Especially for the cavity models, the observability of the peak at systemic velocity depends strongly on the width of the input line since there are many photons near line center that are scattered into the line of sight of observers that look down the cavity. Moreover, our simulations have infinite spectral resolution. In contrast, the finite spectral resolution of real data washes out the detailed structure of the Ly$\alpha$ lines. Both points suggest that looking down a low \ion{H}{I} sightline does not necessarily give rise to a detectable peak in the Ly$\alpha$ spectrum at $x=0$. However, our analysis does suggest that if such a line is detected, then it may be indicative of having a low \ion{H}{I} column density sightline to the galaxy.

\item Dust in general enhances the inclination effect, but the influence differs by an order of magnitude or more between our model families. 

\end{itemize}

A preprint by \citet{Zheng13} appeared recently which also focussed on the escape of Ly$\alpha$ photons from anisotropic gas distributions. While our models share some similarities with theirs, the comparison between their spherical and our shell model extensions are not straightforward since the different geometries have strong effects on the flux distribution. If we quantify the anisotropies by the maximum deviation of the line center optical depth along lines of sight with respect to the mean, as suggested by \cite{Zheng13}, then we find much smaller amplitudes of the inclination effect in comparison to their solid sphere models. For example, in one of our bipolar models, the optical depths along different lines of sight might differ by a factor of $10^3$, while the flux does change only by 10\%. In \cite{Zheng13}, a difference of a factor of 2 in the optical depth can lead to an enhancement in flux as large as a factor of 2. We have confirmed the same enhancement in some of the models of \citet{Zheng13} 
with our code. This suggests that the deviations in the optical depth do not translate directly into an estimate of the flux change for different geometries.

{\bf Acknowledgements} MD thanks the 
the Goettingen Academy of Sciences and Humanities for awarding him a Gauss-Professorship, and the kind hospitality of the 
Institute for Astrophysics in Goettingen.

\begin{appendix}
 
\end{appendix}

\end{document}